\documentclass[manuscript]{aastex}

\usepackage[T1]{fontenc}        

\slugcomment{Accepted for publication in The Astrophysical Journal, March 2015}

\shorttitle{Si and S abundances in flare plasmas}
\shortauthors{Sylwester et al.}

\begin{document}


\title{RESIK SOLAR X-RAY FLARE ELEMENT ABUNDANCES ON A NON-ISOTHERMAL ASSUMPTION}


\author{B. Sylwester\altaffilmark{1},  K. J. H. Phillips\altaffilmark{2}, J. Sylwester\altaffilmark{1}, and A. K\k{e}pa\altaffilmark{1}}
\affil{$^1$ Space Research Center, Polish Academy of Sciences, Kopernika 11, 51-622 Wroc{\l}aw, Poland}
\email{bs@cbk.pan.wroc.pl,js@cbk.pan.wroc.pl,ak@cbk.pan.wroc.pl}
\affil{$^2$ Earth Sciences Department, Natural History Museum, London SW7 5BD, UK}
\email{kennethjhphillips@yahoo.com}



\begin{abstract}

Solar X-ray spectra from the RESIK crystal spectrometer on the {\em CORONAS-F} spacecraft (spectral range $3.3-6.1$~\AA) are analyzed for thirty-three flares using a method to derive abundances of Si, S, Ar, and K, emission lines of which feature prominently in the spectra. For each spectrum, the method first optimizes element abundances then derives the differential emission measure as a function of temperature based on a procedure given by Sylwester et al. and Withbroe. This contrasts with our previous analyses of RESIK spectra in which an isothermal assumption was used. The revised abundances (on a logarithmic scale with $A({\rm H}) = 12$) averaged for all the flares in the analysis are $A({\rm Si}) = 7.53 \pm 0.08$ (previously $7.89 \pm 0.13$), $A({\rm S}) = 6.91 \pm 0.07$ ($7.16 \pm 0.17$), $A({\rm Ar}) = 6.47 \pm 0.08$ ($6.45 \pm 0.07$), and $A({\rm K}) = 5.73 \pm 0.19$ ($5.86 \pm 0.20$), with little evidence for time variations of abundances within the evolution of each flare. Our previous estimates of the Ar and K flare abundances are thus confirmed by this analysis but those for Si and S are reduced. This suggests the flare abundances of Si and Ar are very close to the photospheric abundance or solar proxies, while S is significantly less than photospheric and the K abundance is much higher than photospheric. These estimates differ to some extent from those in which a single enhancement factor applies to elements with first ionization potential less than 10~eV.
\end{abstract}


\section{INTRODUCTION}

In previous work we have used spectra from the X-ray crystal spectrometer RESIK (REntgenovsky Spektrometr s Izognutymi Kristalami: \citet{jsyl05}) to derive element abundances from line fluxes in solar flare and active region spectra. The instrument was mounted on the {\em CORONAS-F} spacecraft and operated from 2001 September to 2003 May, observing numerous flares with {\em GOES} importance levels from B4.6 to X1.5 (for catalog of Level~2 spectra, see http://www.cbk.pan.wroc.pl/experiments/resik/RESIK\_Level2/index.html). This period was one of high solar activity, near the maximum of Cycle~23, which has not been exceeded since. The spectral range of RESIK, $3.3 - 6.1$~\AA, was covered by four channels with incident solar X-rays diffracted by two bent crystals of silicon (Si 111 crystal, $2d = 6.27$~\AA) and two of quartz (quartz $10\bar 10$ crystal, $2d = 8.51$~\AA). Lines due to H- and He-like ions of Si, S, Cl, Ar, and K were included in the range, together with associated dielectronic satellites. The $\sim 20$\% absolute intensity calibration \citep{jsyl05} allows emitting plasma parameters like temperature and emission measure to be deduced with high accuracy, and the presence of nonthermal electrons. Unlike many previous spectrometers, fluorescence due to the crystal material was either eliminated (for the shorter-wavelength channels 1 and 2) or greatly reduced (for channels 3 and 4) by means of on-board pulse-height analyzers, allowing the solar continuum to be observed.

Abundance estimates from most of our previous analyses \citep[see][and references therein]{bsyl13} were made using an isothermal assumption for the emitting plasma with the temperature assumed from the emission ratio of the two {\em GOES} channels. Plots of the estimated fluxes of emission lines from a particular element divided by the {\em GOES} emission measure (EM$_{\rm GOES}$) against {\em GOES} temperature ($T_{\rm GOES}$) were found to follow closely the theoretical functional form of $G(T_e)$ with an assumed element abundance. The departure of the observed points from the $G(T_e)$ curve was generally found to have a distribution close to gaussian, the peak of which defines the estimated abundance and its width the uncertainty. Thus, this procedure was followed for the strong \ion{Ar}{17} lines near 3.95~\AA\ giving an estimated abundance $A({\rm Ar}) = 6.45 \pm 0.07$ (s.d.) \citep{jsyl10b} which we believe to have high reliability. (Abundances are expressed logarithmically on a scale with $A({\rm H}) = 12$.) This value is identical, to within uncertainties, to the best solar proxy estimate, viz. $A({\rm Ar}) = 6.50 \pm 0.10$ (\cite{lod08}: note that Ar spectral lines are not observed in the solar photosphere). For the much weaker \ion{K}{18} and \ion{Cl}{16} lines, the estimated K and Cl abundances \citep{jsyl10a,bsyl11} have greater uncertainties, but the isothermal assumption still appears to hold good. Similar estimates were made for the abundances of Si and S \citep{bsyl13,jsyl12} using the lower-temperature ions \ion{Si}{13}, \ion{Si}{14}, \ion{S}{15}, and \ion{S}{16}, all of which are observed in RESIK's channels 3 and 4, but it was evident from the larger scatter of the points and their temperature dependence that the isothermal assumption was less satisfactory. Our procedure was then to derive the temperature dependence of the differential emission measure (DEM) for each spectrum, as was done in an analysis of non-flaring active region spectra \citep{bsyl10a}. A more recent extensive analysis \citep{bsyl14} (hereafter Paper~I) was made of a single flare on 2002 November~14 in which optimized values of element abundances were first found by a procedure we called AbuOpt, then the DEM derived. The resulting abundance estimates were found to be close to those from the isothermal assumption for Ar and K (Cl lines were ignored as the \ion{Cl}{16} lines are too weak to influence the evaluation of the DEM) but those for Si and S were reduced by factors of 2.2 and 1.8 respectively. The revised S abundance is in fact less than photospheric \citep{asp09,caf11} by a factor of $0.6 \pm 0.1$ for this particular flare.

Our several analyses of RESIK spectra are of interest in the continuing discussion on the nature of the differences between coronal and photospheric abundances and the dependence on first ionization potential (FIP). It has been argued that elements with FIPs less than $\sim 10$~eV (e.g. Si, K) have enhanced abundances in coronal plasma (by factors of up to $\sim 4$ according to \cite{fel92b}) relative to photospheric abundances, while those with FIPs greater than 10~eV (e.g. Cl, Ar) have comparable coronal and photospheric abundances. In this paper, we extend our work using the AbuOpt approach to 2418 spectra obtained during thirty-three flares and thus generalize and refine the results obtained in Paper~I. The data and analysis are described in Section~2, and discussion of the resulting abundance estimates and their relation to the FIP are given in Section~3.

\section{OBSERVATIONS AND ANALYSIS}

Data for the thirty-three flares used in this analysis are given in Table~\ref{flare_list}. The dates and times of peak X-ray are defined in the IAU-approved flare-naming convention, together with {\em GOES} importance and flare location. The numbers of spectra for each flare used in the analysis are also given. The analysis described here is based on a calculational procedure (the ``Withbroe--Sylwester'' method) to derive the differential emission measure DEM $= \varphi(T_e) = N_e^2 dV / dT_e$ ($V = $ emitting volume, $N_e = $ electron density), which has been developed over several years starting with work by \cite{with75} and \cite{jsyl80}.

\begin{figure}
\epsscale{.60}
\plotone{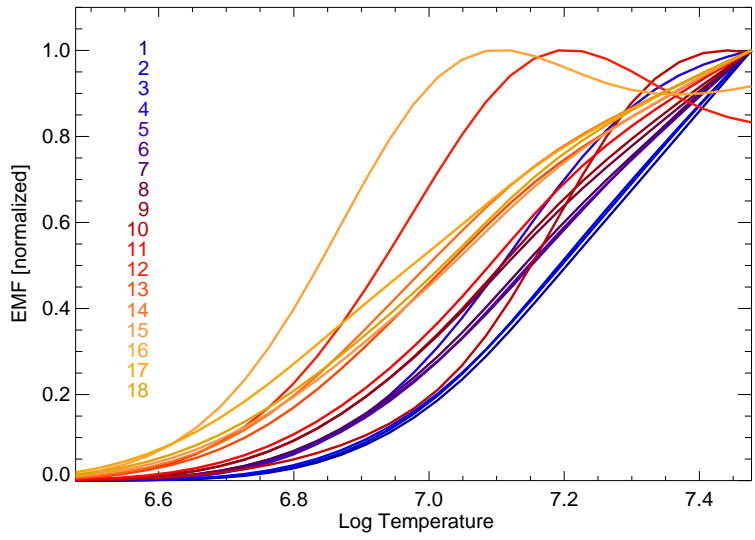}
\caption{Normalized emissivity functions (emission per unit volume emission measure) plotted against log~$T_e$ of the 18 channels listed in Table~\ref{wavelength_bands}. The colored curves are identified by the colored numbers in the legend which refer to the wavelength band numbers in Table~\ref{wavelength_bands}.
\label{emissivities}}
\end{figure}

The input data consist of observed fluxes $F_i$ in 18 narrow spectral bands over the RESIK wavelength range which are defined in Table~\ref{wavelength_bands}. These bands, which are different from those used in Paper~I, include both lines and continuum and continuum alone. For flare plasmas which are in general multi-thermal, the fluxes in each spectral interval are

\begin{equation}
F_i = A_i \int^{\infty}_{T_e=0} f_i(T_e) \varphi(T_e) dT_e
\end{equation}

\noindent where the abundance of an element contributing to the flux $F_i$ is $A_i$. The emissivity functions $f_i$  (i.e. observed fluxes per unit volume emission measure) are functions of $T_e$. They were evaluated with the {\sc chianti} (v.~7.0) database and code \citep{der97,lan12}. Normalized values are plotted in Figure~\ref{emissivities}, showing that there is a reasonably uniform temperature distribution of emissivities above log~$T_e \sim 6.9$ ($T_e$ in K), with lower temperatures principally defined by emission lines of He-like Si (\ion{Si}{13}) and dielectronic satellites of \ion{Si}{12} which strongly feature in RESIK channel~4 flare spectra and were discussed by \cite{phi06}. This is important for an accurate evaluation of the DEM function $\varphi$ as a function of $T_e$ from the inversion of the integral in Equation~1.

As described in Paper~I, optimized abundances of elements are first derived from the observed fluxes for each of the spectra taken during the 33 flares listed in Table~\ref{flare_list}. We then ran the Withbroe-Sylwester DEM code which has been developed based on the method of \cite{with75} and which is described by \cite{jsyl80}. In essence, a first-order solution to $\varphi$ in a particular temperature interval with observed flux $F_i$ is derived as a mean value, $\langle \varphi \rangle$, over this interval; a weighting function to this solution is then applied that corrects the resulting flux $F'_i$ to the observed value $F_i$. Revised values of $\varphi$ are estimated iteratively until a complete solution to $\varphi(T_e)$ that leads to flux estimates within observational uncertainties of the observed fluxes $F_i$. Thus, the method does not assume a functional form for $\varphi(T_e)$ and is in that sense {\em ab initio}.

The inversion of Equation~(1) to derive $\varphi(T_e)$ is subject to uncertainties in the observational data although, as discussed in our previous paper, we checked that solutions for $\varphi(T_e)$ and their uncertainties satisfied the input data to within observational uncertainties. There are also uncertainties to consider in the atomic data. These have been discussed by \cite{phi75} and \cite{crabro76} and more recently \cite{gue13}. For the emission lines included in the RESIK wavelength bands (Table~\ref{wavelength_bands}), the most important uncertainties include ionization fractions (taken from a default database which is identical to the data of \cite{bry09}) and temperature-averaged excitation rate coefficients (from {\sc chianti} based on distorted wave calculations of \cite{zha87}). The spectral lines in the RESIK bands are emitted by H-like or He-like ions or are Li-like ion dielectronic satellites. The ionization fractions appropriate to these lines are extremely close to those from much earlier calculations (e.g., \cite{maz98}) and do not appear to be subject to any significant revision. The excitation rates for these lines have been calculated more recently but show little difference from the earlier distorted-wave data. Thus, \cite{agg10} have used the close-coupling $R$-matrix code to calculate excitation rates for \ion{Si}{13} lines which are only a few percent different from the \cite{zha87}. The errors in the derivation of $\varphi(T_e)$ arising from atomic data are therefore likely to be small.

Over the operational lifetime of RESIK, continual adjustments were made to the instrument's pulse-height analyzer settings to minimize the crystal fluorescence background. Eventually, this background was entirely eliminated for the short-wavelength channels~1 and 2 (range $3.40-4.27$~\AA\ including \ion{K}{18}, \ion{Ar}{17}, \ion{Ar}{18}, and some \ion{S}{15} lines) and greatly reduced for the long-wavelength channels 3 and 4 (range $4.35-6.05$~\AA\ including \ion{Si}{13}, \ion{Si}{14}, and \ion{S}{15} lines as well as \ion{Si}{12} dielectronic satellites). This optimization was achieved over the period 2002 December~24 (01:00~UT) to 2003 April 5 (00:45~UT). The 33 flares analyzed here occurred during this time range. In Table~\ref{flare_list}, values of the estimated abundances of Si, S, K, and Ar averaged over spectra considered by us to be of good quality are given. They were obtained using the procedure described for each flare with uncertainties which are the standard deviations in the estimated abundances for the number of spectra analyzed. At the foot of the table are means of the 33 estimated abundances with their standard deviations and these values divided by photospheric abundances. Also at the bottom of the table are photospheric abundances from \cite{asp09} or, in the case of Ar, solar proxies (based on various non-solar sources: see \cite{lod08}). Also given are estimated abundances from analysis of RESIK spectra using an isothermal approximation which, as described below, are likely to be less certain for Si and S.

The estimated abundances obtained here for Si, S, and Ar are very similar to the values obtained in the analysis of the X-ray M1 class flare on 2002 November~14 (Paper~I), and for Ar and K they are also very similar to estimates from our isothermal analyses of RESIK flares \citep{jsyl10a,jsyl10b}. For Si and S, our present estimates are lower than those from isothermal analyses \citep{bsyl13,jsyl12} by factors of approximately 2. The difference can be attributed to the non-isothermal character of the flare plasma in which the {\em GOES} temperature is not a good representation of the flare's temperature structure but it is so for the case of K and Ar, a point which we discuss further below.

\begin{figure}
\epsscale{.60}
\plotone{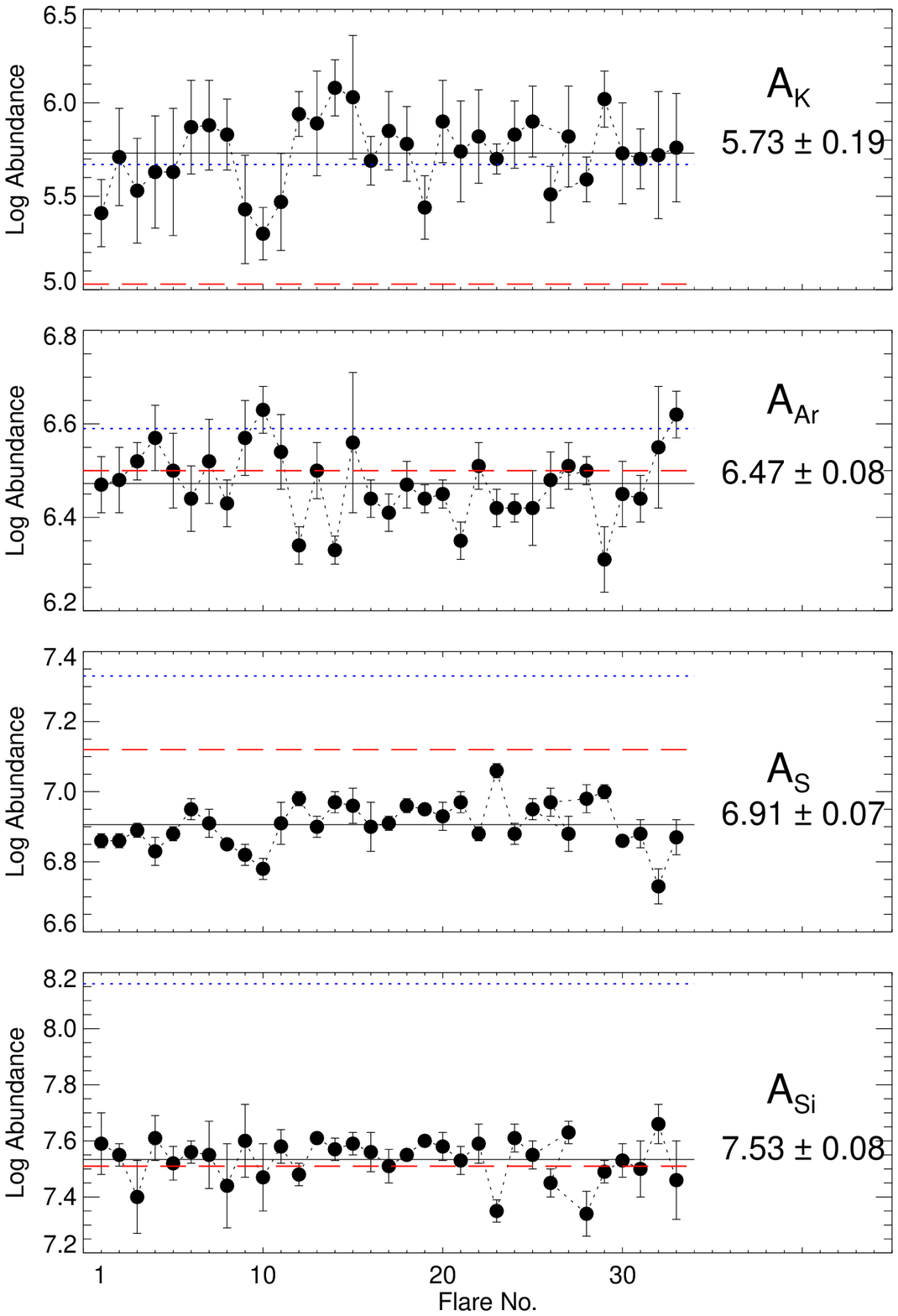}
\caption{Mean abundance estimates for each of the 33 flares listed in Table~\ref{flare_list} for K, Ar, S, and Si. The red dashed horizontal lines indicate photospheric abundance estimates from \cite{asp09} (Si, S, K) or solar proxies from \cite{lod08} (Ar). Blue dotted lines are coronal abundance estimates from \cite{fel92a} (Si, S, Ar) and those specified in the {\sc chianti} database ``coronal'' abundance file (K). Black horizontal lines indicate abundance values averaged over all 33 flares.
\label{abu_flare_number}}
\end{figure}

Figure~\ref{abu_flare_number} is a plot of the estimated abundances against flare number as given in Table~\ref{flare_list} with error bars given by uncertainties estimated as above. Horizontal lines indicate photospheric abundances (red dashed lines, from \cite{asp09}, with Ar from \cite{lod08}) and coronal abundances (blue dotted lines), from \cite{fel00}. There is no apparent relation of the abundances with {\em GOES} X-ray class (the total range is from B9.9 to X1.5) nor on the disposition of the flare on the solar disk (the flares analyzed include those near disk center as well as near the limb).

\begin{deluxetable}{clcc}
\tabletypesize{\scriptsize} \tablecaption{Wavelength bands used in DEM analysis \label{wavelength_bands}}
\tablewidth{0pt}
\tablehead{\colhead{No.}& \colhead{Wavelength range (\AA)} & \colhead{RESIK Channel} & \colhead{Emission features included} \\}
\startdata
1 & 3.40 - 3.50 & 1 & continuum \\
2 & 3.50 - 3.60 & 1 & K XVIII triplet + sats. \\
3 & 3.60 - 3.80 & 1 & continuum \\
4 & 3.915 - 4.025 & 2 & Ar XVII triplet + sats. \\
5 & 4.11 - 4.17 & 2 & continuum \\
6 & 4.17 - 4.21 & 2 & S XV $1s^2-1s4p$ sats. \\
7 & 4.21 - 4.25 & 2 & continuum \\
8 & 4.36 - 4.42 & 3 & S XV d3 sats. \\
9 & 4.42 - 4.68 & 3 & continuum \\
10 & 4.70 - 4.75 & 3 & S XVI Ly$\alpha$ \\
11 & 4.75 - 4.80 & 3 & S XV sats to \ion{S}{16} Ly$\alpha$ \\
12 & 5.00 - 5.15 & 4 & S XV triplet + sats. \\
13 & 5.25 - 5.32 & 4 & Si XIII $1s^2-1s5p$ \\
14 & 5.37 - 5.48 & 4 & Si XIII $1s^2-1s4p$ + $5p$ sat. \\
15 & 5.48 - 5.62 & 4 & Si XII $4p$ sats. \\
16 & 5.645 - 5.71 & 4 & Si XIII $1s^2 - 1s3p$ \\
17 & 5.77 - 5.86 & 4 & Si XII $3p$ sat. \\
18 & 5.92 - 5.97 & 4 & continuum \\
\enddata
\end{deluxetable}


\begin{figure}
\epsscale{.62}
\plotone{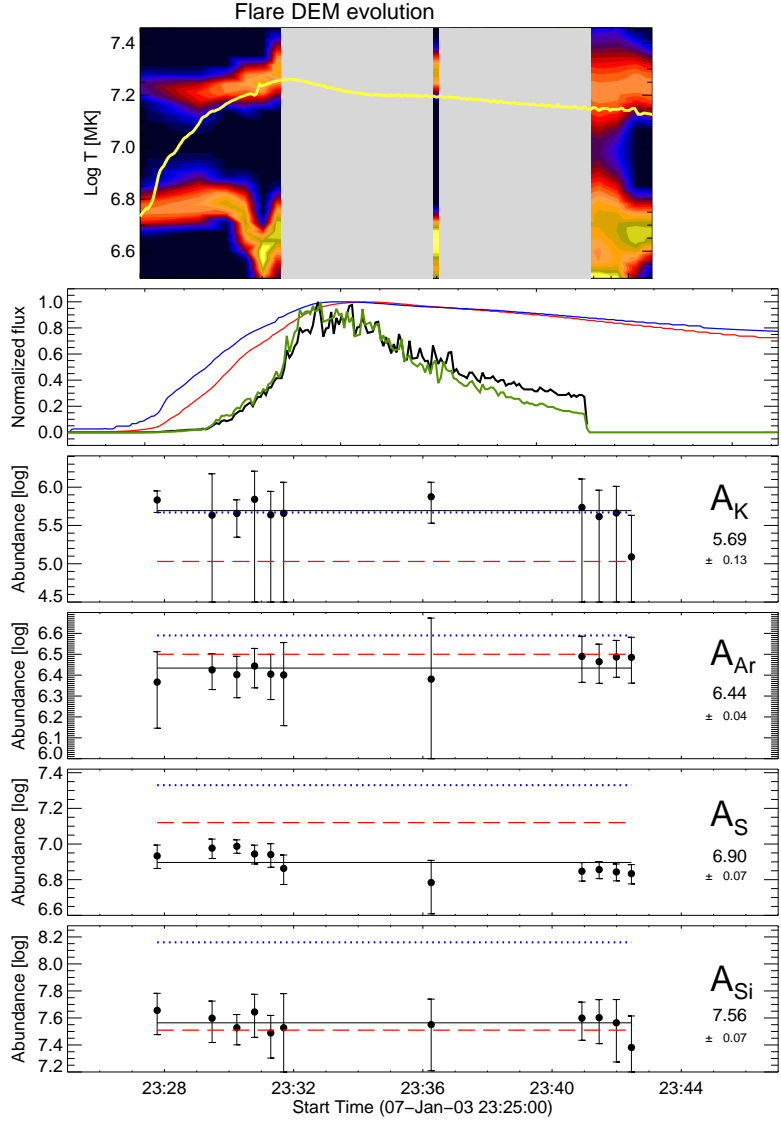}
\caption{Top panel: Flare differential emission measure (DEM) evolution for the 2003 January~7 event (no. 16 in Table~\ref{flare_list}). The horizontal time axis is identical to that on the bottom scale. Gray stripes show missing data periods, while the yellow line is the average temperature from the \emph{GOES} channels flux ratio. Second panel from top: Normalized {\em GOES} flux in $0.5-4$~\AA\ (blue line) and $1-8$~\AA\ (red line) bands, with impulsive rise over the 23:28--23:32~UT period. The flare's impulsive burst is evident in {\em RHESSI} light curves (3--6~keV in black; 25--50~keV in green). Lower four panels: Abundance estimates (logarithmic scale, H=12) for K, Ar, S, and Si from spectra integrated over eleven time intervals (mean values with standard deviations given in the legend in the right of each panel). Horizontal colored lines have the same meaning as in Figure~\ref{abu_flare_number}.
\label{abu_time_2003-Jan-07}}
\end{figure}

\begin{figure}
\epsscale{.65}
\plotone{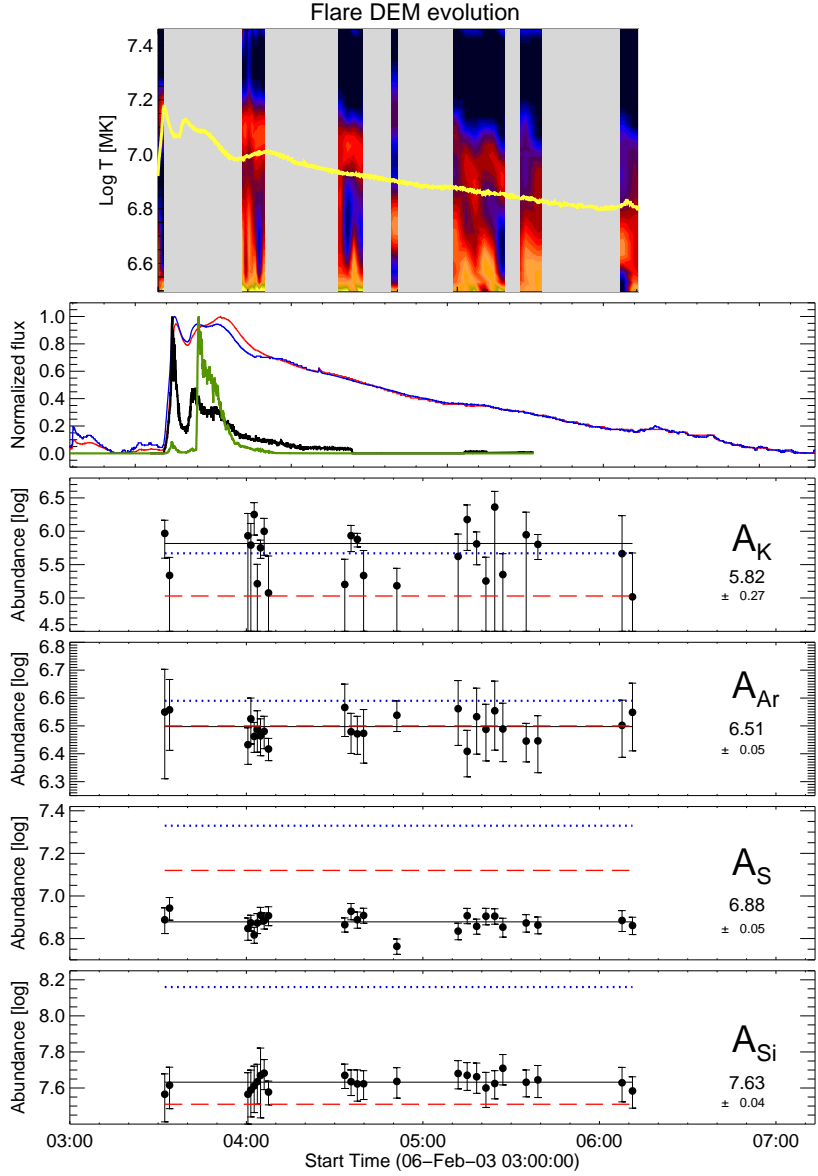}
\caption{Top panel: Flare DEM evolution for the 2003 February~6 event (no. 26 in Table~\ref{flare_list}). Gray stripes and yellow line indicate missing data periods and average {\em GOES} temperature respectively. Second panel from top: Normalized {\em GOES} flux in $0.5-4$~\AA\ (blue line) and $1-8$~\AA\ (red line) bands, with {\em RHESSI} light curves (3--6~keV in black; 12--25~keV in green). Lower four panels: Abundance estimates (logarithmic scale, H=12) for K, Ar, S, and Si from spectra integrated over 24 time intervals (mean values with standard deviations in the legend). Horizontal colored lines have the same meaning as in Figure~\ref{abu_flare_number}.
\label{abu_time_2003-Feb-06}}
\end{figure}

\cite{fel90b} cite an example of an impulsively heated flare observed in 1973 by the {\em Skylab} extreme ultraviolet spectroheliograph, in which the O/Mg abundance ratio was close to the photospheric abundance
ratio, while on the standard FIP picture the low-FIP Mg would be expected to be enhanced in a coronal plasma but not the high-FIP O. This is attributed by \cite{fel00} to photospheric material being suddenly heated to coronal temperatures in the flare's impulsive stage. We investigated whether such effects occur for any of the flares given in Table~\ref{flare_list}, although only two show a sudden impulsive rise in X-ray emission as indicated by {\em GOES} light curves and which are reasonably well observed by RESIK. Figure~\ref{abu_time_2003-Jan-07} (lower five panels) shows the {\em GOES} light curves and abundance estimates of Si, S, Ar, and K from a total of 112 spectra integrated over eleven time intervals (to improve count statistics) during the flare of 2003 January~7. Over a four-minute period (23:28--23:32~UT), there is a steep rise of the {\em GOES} X-ray emission but there is little response in the abundances apart from S which shows a barely significant abundance increase over this period by a factor of approximately 2 compared with later times when the X-ray emission shows a slower decline after its maximum. The K abundance is not well determined for this flare for most individual spectra, but its mean abundance is clearly higher than the photospheric value. The abundances of Si and Ar show no particular trend over the entire flare. Figure~\ref{abu_time_2003-Feb-06} is the corresponding plot for the flare of 2003 February~6 (179 spectra integrated over 24 time intervals), with very similar results although only two RESIK mean spectra (averaged over four and five data gathering intervals) were obtained during the impulsive rise (03:31~UT) of this flare. The impulsiveness of the rise of both these flares does not quite match that of the 1973 flare seen by {\em Skylab} (a total of three minutes from flare onset to attaining temperatures of several $10^5$~K), so the question of varying abundances for highly impulsive flares is arguably not completely settled by RESIK data but equally there is no convincing evidence for sharp increases in, say, the low-FIP K abundance. In the top panels of Figures~\ref{abu_time_2003-Jan-07} and \ref{abu_time_2003-Feb-06}, the time evolution of the differential emission measure for the optimized abundances is shown as a color contour plot. Unfortunately, several missing data periods are present for each of these flares, but there is a reasonably clear tendency for a two-component temperature structure over most of the flare development, as was found in the case of the 2002 November~14 flare analyzed in Paper~I and has been noted by other investigators (e.g. \cite{mct99}).

\section{DISCUSSION AND CONCLUSIONS}

This analysis of $3.3-6.1$~\AA\ spectra from the RESIK X-ray crystal spectrometer, taken during 33 flares between 2002 and 2003, has used a different approach to determining abundances from that used in most of our previous work. In particular an optimization of abundances has been carried out with determination of differential emission measure (DEM) for individual spectra. The elements concerned, Si, S, Ar, and K, have readily distinguishable lines, the fluxes of which can be determined with high precision using the intensity calibration of the instrument \citep{jsyl05}, and the flares included in the analysis occurred during a period when the instrument's pulse height analyzers were adjusted to optimum values. Compared with our previous isothermal analyses \citep{bsyl13,jsyl12}, the present analysis gives a silicon abundance which is a factor 2.1 lower and a sulfur abundance which is a factor 1.8 lower. Compared with our isothermal analyses for Ar and K \citep{jsyl10a,jsyl10b}, the present estimate is only slightly (5\%) lower for Ar and 40\% lower for K (based on the much weaker \ion{K}{18} lines), both differences being within estimated uncertainties.

The reliability of the isothermal analysis depends on the closeness of observed values of line flux divided by {\em GOES} emission measure plotted at values of the {\em GOES} temperature to the calculated $G(T_e)$ curve. In the case of Ar particularly and K, the agreement of observed points with the calculated curve was very satisfactory but was less so for Si and S. This appears to be due to the fact that the temperature from the ratio of the two {\em GOES} channels is a good description of the temperature characterizing the \ion{Ar}{17} and \ion{K}{18} lines which were mainly used in the Ar and K abundance determinations. The {\em GOES} temperature is however significantly higher than the temperatures characterizing the \ion{S}{15}, \ion{S}{16}, \ion{Si}{13}, and \ion{Si}{14} lines which appears to explain the departure of points for these lines from the $G(T_e)$ curve and why the abundances are better determined when a differential emission method is used.

In summary, this analysis has resulted in the following abundance estimates: $A({\rm Si}) = 7.53 \pm 0.08$, $A({\rm S}) = 6.91 \pm 0.07$, $A({\rm Ar}) = 6.47 \pm 0.08$, and $A({\rm K}) = 5.73 \pm 0.19$. These Si and S abundance estimates are to be preferred to those from our previous analyses \citep{bsyl13,jsyl12}. The present estimates are very close to those obtained previously for Ar and K. There is little evidence of abundance variations with flare time evolution apart from a hint of a rise in the S abundance  during the fast rise in the 2003 January~7 flare. Flare-to-flare variations, such as were observed for Ca \citep{jsyl98}, seem to be ruled out for Si, S, and Ar but $\pm 50$\% variations are not ruled out for K in view of the larger uncertainty in its abundance estimate.

Our result that the S abundance is only $0.6 \pm 0.1$ times photospheric is identical to that of \cite{vec81} who analyzed spectra from the graphite crystal spectrometer on the {\em Orbiting Solar Observatory-8} spacecraft. The advantage of using this low-$Z$ material as a diffracting crystal is that fluorescence is eliminated and so the solar continuum can be distinguished. \cite{vec81} used the continuum to obtain temperature and the observed \ion{S}{15} line fluxes to obtain the S abundance, found to be $6.91^{+.13}_{-.19}$, the same as the value found here. Our estimate for the Si abundance is also near to the value obtained by \cite{vec81} from \ion{Si}{14} lines, $7.62^{+0.19}_{-0.34}$ and both differ only slightly from the photospheric abundance, $7.51 \pm 0.03$. Abundance estimates for these elements are also available from  the X-ray Solar Monitor, a broad-band instrument on the Indian spacecraft {\em Chandrayaan-1} \citep{nar14}. The energy resolution (200~eV at 5.9~keV) is adequate to resolve spectral features due to Si and S. Although only small flares were observed (the spacecraft operated during the deep solar minimum of 2009), analysis of a C3 flare gave $A({\rm Si}) = 7.47 \pm 0.10$ and $A({\rm S}) = 7.16 \pm 0.17$, the Si estimate being similar to ours and that of \cite{vec81} and the S estimate marginally higher but just within uncertainty limits.

The ratio of Si/S abundances found here from our sample of flares is $4.2^{+1.7}_{-1.2}$, which is of some interest for recent investigations of this abundance ratio in non-flaring active regions using spectra from the Extreme ultraviolet Imaging Spectrometer (EIS) on the {\em Hinode} spacecraft. This was found to vary over the range $2.5-4.1$ by \cite{bro11} and $2.5-3$ by \cite{bak13}. The heating and consequent fractionation of elements may be different in active regions from flares, but it appears that our measurements are consistent with these observations. We also note that the agreement with our result arises not because of the enhancement of the low-FIP element Si but rather the {\em reduced} S abundance compared with photospheric.

Our abundance estimates differ in some respects from those given in several discussions \citep[see, e.g.,][]{fel92a,fel00,flu99} relating deviations to the first ionization potential (FIP) of the element, in particular whether the FIP is less or greater than 10~eV (low-FIP or high-FIP). Thus, \cite{fel00} state that low-FIP elements are enriched in the corona by factors of between 1 and 4 for active regions, tending to increase with active region age. However, according to the estimates given here, Si, a low-FIP element, has a flare abundance which is within uncertainties identical to the photospheric abundance.  For S, an element with ``intermediate'' FIP (10.36~eV), our flare abundance estimate is only $0.6 \pm 0.1$ of the photospheric abundance. The consistency of our estimates from flare to flare with a spread of active region ages does not indicate any age dependence. On the other hand, the estimated abundance of K, with very low FIP (4.34~eV), is a factor $5 \pm 2$ higher than photospheric and that of the high-FIP argon is close to photospheric, in agreement with \cite{fel00}. We note that the photospheric abundances of Si, S, and K have small uncertainties \citep{asp09}, as do the proxy values of Ar \citep{lod08}.

A comparison of our results with FIP models is obviously relevant but unfortunately most models (those up to 1998 are summarized by \cite{hen98}) do not give quantitative predictions as to the abundances of elements in coronal plasma compared with photospheric. The model developed by \cite{lam12} (see also references therein) uses the idea of a ponderomotive force generated by Alfv\'{e}n waves passing through the solar atmosphere to accelerate low-FIP elements into the corona from the fractionation region, taken to be at the top of the solar chromosphere where low-FIP elements are ionized. Our results require an ``inverse'' FIP bias for S but a positive FIP bias for the low-FIP element K, and no enhancement for Si and Ar. A sample calculation by \cite{lam12} for a loop length of $100\,000$~km predicts enhancements for K but none for Ar, as is observed for our flares, but also small enhancements for Si and S which are not observed. Possibly varying model parameters including a more realistic loop length for flares will match our results better.

\acknowledgments

We acknowledge financial support from the Polish National Science Centre grant number 2011/01/B/ST9/05861 and the European Commissions grant FP7/2007-2013: eHEROES, Project No. 284461. The {\sc chianti} atomic database and code is a collaborative project involving George Mason University, University of Michigan (USA), and University of Cambridge (UK).

{\em Facilities:} \facility{GOES}, \facility{CORONAS/RESIK}, \facility{RHESSI}

\bibliographystyle{apj}

\bibliography{RESIK}

\newpage
\begin{deluxetable}{clcccccccc}
\tabletypesize{\scriptsize} \tablecaption{RESIK Flares in this Analysis \label{flare_list}}
\tablewidth{0pt}
\tablehead{\colhead{Flare}& \colhead{Flare notation} & \colhead{{\em GOES}} &\colhead{Location} & \colhead{No. of}& \colhead{$A({\rm Si})^a$}  & \colhead{$A({\rm S})$} & \colhead{$A({\rm Ar})$} & \colhead{$A({\rm K})$} \\
\colhead{no.}&&\colhead{class}&& \colhead{spectra} \\}
\startdata

01 & SOL2002-12-25T05:46 & C4.8 & S15W89 & 48 & $7.59 \pm 0.11$ & $6.86 \pm 0.02$ & $6.47 \pm 0.06$ & $5.41 \pm 0.18$ \\
02 & SOL2002-12-25T06:02 & C4.0 & S15W89 & 67 & $7.55 \pm 0.04$ & $6.86 \pm 0.02$ & $6.48 \pm 0.07$ & $5.71 \pm 0.26$ \\
03 & SOL2002-12-25T12:07 & C3.5 & S29W90 & 63 & $7.40 \pm 0.13$ & $6.89 \pm 0.02$ & $6.52 \pm 0.04$ & $5.53 \pm 0.28$ \\
04 & SOL2002-12-25T13:55 & C1.1 & S29W90 & 19 & $7.61 \pm 0.08$ & $6.83 \pm 0.04$ & $6.57 \pm 0.07$ & $5.63 \pm 0.30$ \\
05 & SOL2002-12-25T18:09 & C2.9 & S30W90 & 82 & $7.52 \pm 0.06$ & $6.88 \pm 0.02$ & $6.50 \pm 0.08$ & $5.63 \pm 0.34$ \\
06 & SOL2002-12-25T21:32 & C1.7 & S08W44 & 22 & $7.56 \pm 0.04$ & $6.95 \pm 0.03$ & $6.44 \pm 0.07$ & $5.87 \pm 0.25$ \\
07 & SOL2002-12-26T02:01 & C1.4 & S08W46 & 29 & $7.55 \pm 0.12$ & $6.91 \pm 0.04$ & $6.52 \pm 0.09$ & $5.88 \pm 0.24$ \\
08 & SOL2002-12-26T08:35 & C1.9 & S00E43 & 53 & $7.44 \pm 0.15$ & $6.85 \pm 0.01$ & $6.43 \pm 0.05$ & $5.83 \pm 0.19$ \\
09 & SOL2002-12-28T23:14 & C1.4 & N11E24 & 19 & $7.60 \pm 0.13$ & $6.82 \pm 0.03$ & $6.57 \pm 0.08$ & $5.43 \pm 0.29$ \\
10 & SOL2002-12-29T02:05 & B9.9 & N15W43 & 29 & $7.47 \pm 0.12$ & $6.78 \pm 0.03$ & $6.63 \pm 0.05$ & $5.30 \pm 0.14$ \\
11 & SOL2003-01-07T04:35 & C1.6 & S10E90 & 26 & $7.58 \pm 0.06$ & $6.91 \pm 0.06$ & $6.54 \pm 0.08$ & $5.47 \pm 0.26$ \\
12 & SOL2003-01-07T07:50 & M1.0 & S24E08 & 237 & $7.48 \pm 0.04$ & $6.98 \pm 0.02$ & $6.34 \pm 0.04$ & $5.94 \pm 0.12$ \\
13 & SOL2003-01-07T08:33 & C7.7 & S12E90 & 9 & $7.61 \pm 0.02$ & $6.90 \pm 0.03$ & $6.50 \pm 0.06$ & $5.89 \pm 0.28$ \\
14 & SOL2003-01-07T11:12 & C2.9 & S29E90 & 59 & $7.57 \pm 0.04$ & $6.97 \pm 0.03$ & $6.33 \pm 0.03$ & $6.08 \pm 0.15$ \\
15 & SOL2003-01-07T12:35 & C1.3 & S15E90 & 9 & $7.59 \pm 0.04$ & $6.96 \pm 0.05$ & $6.56 \pm 0.15$ & $6.03 \pm 0.33$ \\
16 & SOL2003-01-07T23:33$^b$ & M4.9 & S11E89 & 112 & $7.56 \pm 0.07$ & $6.90 \pm 0.07$ & $6.44 \pm 0.04$ & $5.69 \pm 0.13$ \\
17 & SOL2003-01-09T01:39 & C9.8 & S09W25 & 178 & $7.51 \pm 0.06$ & $6.91 \pm 0.02$ & $6.41 \pm 0.04$ & $5.85 \pm 0.21$ \\
18 & SOL2003-01-09T17:26 & C2.4 & S11E57 & 39 & $7.55 \pm 0.02$ & $6.96 \pm 0.02$ & $6.47 \pm 0.05$ & $5.78 \pm 0.20$ \\
19 & SOL2003-01-21T02:28 & C8.1 & N14E09 & 69 & $7.60 \pm 0.02$ & $6.95 \pm 0.01$ & $6.44 \pm 0.03$ & $5.44 \pm 0.17$ \\
20 & SOL2003-01-21T02:45 & C2.8 & N14E09 & 25 & $7.58 \pm 0.05$ & $6.93 \pm 0.04$ & $6.45 \pm 0.03$ & $5.90 \pm 0.22$ \\
21 & SOL2003-01-21T15:26 & M1.9 & S07E90 & 181 & $7.53 \pm 0.05$ & $6.97 \pm 0.03$ & $6.35 \pm 0.04$ & $5.74 \pm 0.27$ \\
22 & SOL2003-01-25T18:55 & C4.4 & N13W27 & 52 & $7.59 \pm 0.07$ & $6.88 \pm 0.02$ & $6.51 \pm 0.05$ & $5.82 \pm 0.25$ \\
23 & SOL2003-01-27T22:19 & C2.4 & S17W24 & 85 & $7.35 \pm 0.04$ & $7.06 \pm 0.02$ & $6.42 \pm 0.04$ & $5.70 \pm 0.08$ \\
24 & SOL2003-02-01T09:05 & M1.2 & S15E90 & 133 & $7.61 \pm 0.05$ & $6.88 \pm 0.03$ & $6.42 \pm 0.03$ & $5.83 \pm 0.18$ \\
25 & SOL2003-02-06T02:12 & C3.4 & S16E55 & 34 & $7.55 \pm 0.05$ & $6.95 \pm 0.03$ & $6.42 \pm 0.08$ & $5.90 \pm 0.19$ \\
26 & SOL2003-02-06T03:49$^b$ & M1.2 & N19E63 & 171 & $7.63 \pm 0.04$ & $6.88 \pm 0.05$ & $6.51 \pm 0.05$ & $5.82 \pm 0.27$ \\
27 & SOL2003-02-14T02:12 & C5.4 & N12W88 & 44 & $7.45 \pm 0.05$ & $6.97 \pm 0.04$ & $6.48 \pm 0.06$ & $5.51 \pm 0.15$ \\
28 & SOL2003-02-14T05:26 & C5.6 & N11W85 & 64 & $7.34 \pm 0.08$ & $6.98 \pm 0.04$ & $6.50 \pm 0.03$ & $5.59 \pm 0.12$ \\
29 & SOL2003-02-15T08:10 & C4.5 & S10W89 & 299 & $7.49 \pm 0.04$ & $7.00 \pm 0.02$ & $6.31 \pm 0.07$ & $6.02 \pm 0.15$ \\
30 & SOL2003-02-21T19:50 & C4.3 & N15E01 & 32 & $7.53 \pm 0.06$ & $6.86 \pm 0.01$ & $6.45 \pm 0.07$ & $5.73 \pm 0.27$ \\
31 & SOL2003-02-22T09:29 & C5.8 & N16W05 & 24 & $7.50 \pm 0.10$ & $6.88 \pm 0.04$ & $6.44 \pm 0.05$ & $5.70 \pm 0.16$ \\
32 & SOL2003-02-22T12:20 & C1.7 & N16W07 & 73 & $7.66 \pm 0.07$ & $6.73 \pm 0.05$ & $6.55 \pm 0.13$ & $5.72 \pm 0.34$ \\
33 & SOL2003-03-17T19:05 & X1.5 & S14W38 & 32 & $7.46 \pm 0.14$ & $6.87 \pm 0.05$ & $6.62 \pm 0.05$ & $5.76 \pm 0.29$\\
Mean values$^c$ & & & & & $7.53 \pm 0.08$ & $6.91 \pm 0.07$ & $6.47 \pm 0.08$ & $5.73 \pm 0.19$ \\
Enhancement  &over photospheric& & & & $1.05 \pm 0.2$ & $0.62 \pm 0.1$ & $0.93 \pm 0.15$ & $5.0 \pm 2$\\
Isothermal &analysis$^d$& & & & $7.89 \pm 0.13$ & $7.16 \pm 0.17$ & $6.45 \pm 0.07$ & $5.86 \pm 0.20$\\
Photospheric$^e$ & & & & & $7.51 \pm 0.03$ & $7.12 \pm 0.03$ & $6.50 \pm 0.10$ & $5.03 \pm 0.09$   \\
FIP (eV) & & & & & 8.15 & 10.36 & 15.76 & 4.34 \\
\enddata
\tablenotetext{a} {Abundances are expressed on a logarithmic scale with $A({\rm H}) = 12$.}
\tablenotetext{b} {Values of irradiances in the 18 spectral bands (Table~\ref{wavelength_bands}) for flares nos. 16 and 26 are given as a file attached to this article. }
\tablenotetext{c} {Mean values of all 33 flares.}
\tablenotetext{d} {See text for references to isothermal analyses of RESIK spectra.}
\tablenotetext{e} {Photospheric abundances from \cite{asp09}; Ar from solar proxies \citep{lod08}.}

\end{deluxetable}

\end{document}